# ELECTRON BEAM DYNAMICS IN 4GLS

P. H. Williams, G. Hirst, B. D. Muratori, H. L. Owen, S. L. Smith
STFC Daresbury Laboratory & Cockcroft Institute, Warrington, WA4 4AD, UK

*Abstract*

Studies of the electron beam dynamics for the 4GLS design are presented. 4GLS will provide three different electron bunch trains to a variety of user synchrotron sources. The 1 kHz XUV-FEL and 100 mA high-current ERL branches share a common 540 MeV linac, whilst the 13 MHz IR-FEL must be well-synchronised to them. An overview of recent developments in the optics design of the facility is given.

## THE 4GLS FACILITY

The 4GLS facility, proposed to be constructed at Daresbury Laboratory in the UK, will consist of three inter-related accelerator systems each driving a free-electron laser; these lasers will deliver short-pulse output in the terahertz, infra-red, VUV and soft X-ray portions of the electromagnetic spectrum with pulse lengths as short as 50 fs. In combination with spontaneous output from undulators, this multi-source, multi-user facility will enable the study of real-time molecular processes on the femtosecond timescale. A conceptual design has been produced [1,2], and the scientific motivations for the project are described elsewhere [3]. The overall lattice design has been published in an EPAC 2006 paper [4], we present here a selection of major design developments since that time. Figure 1 shows a schematic layout of 4GLS. The high-current ERL is the outer loop of the facility and contains five insertion devices. The last of these utilises fully-compressed ~100 fs, 77 pC bunches to drive a regenerative-amplifier VUV-FEL. The 1nC XUV bunches are propagated on the opposing RF phase in the main linac, then separated with a beam spreader and solenoid, passed through 180° FODO arc tilted out of plane and transported to two XUV-FEL's at ~2m vertical displacement from the linac-ERL plane.

## OPPOSING-PHASE ACCELERATION & COMPRESSION

The most challenging part of the 4GLS project is the design and construction of an energy recovery linac that will deliver 100 mA of average beam current through five insertion device straights with small transverse emittance and short bunch lengths. 77 pC bunches from a ~750 kV DC photo-injector are pre-accelerated to 10 MeV by two five-cavity superconducting RF modules [5]. They then make an energy-recovery pass of the main linac and insertion device arc at 550 MeV and are dumped at 10 MeV. The main linac also accelerates in single pass configuration at 1 kHz, 1 nC bunches that drive two seeded XUV-FEL's [6]: 540 MeV acceleration is also required by these bunches. The final bunch parameters for all three accelerator channels are summarised in Table 1.

Since the main linac accelerates two types of bunch, we must keep them apart so that they do not interfere. Our proposed solution to keep the two bunch types apart is a novel scheme whereby the XUV and high-current loop bunches are accelerated on opposing phases of the main linac RF. The bunches thereby receive opposite signed energy chirps. The subsequent compression scheme is then arranged to be chicane-like in the XUV-FEL branch and arc-like in the high-current loop.

For further details see the EPAC 2006 paper [4].

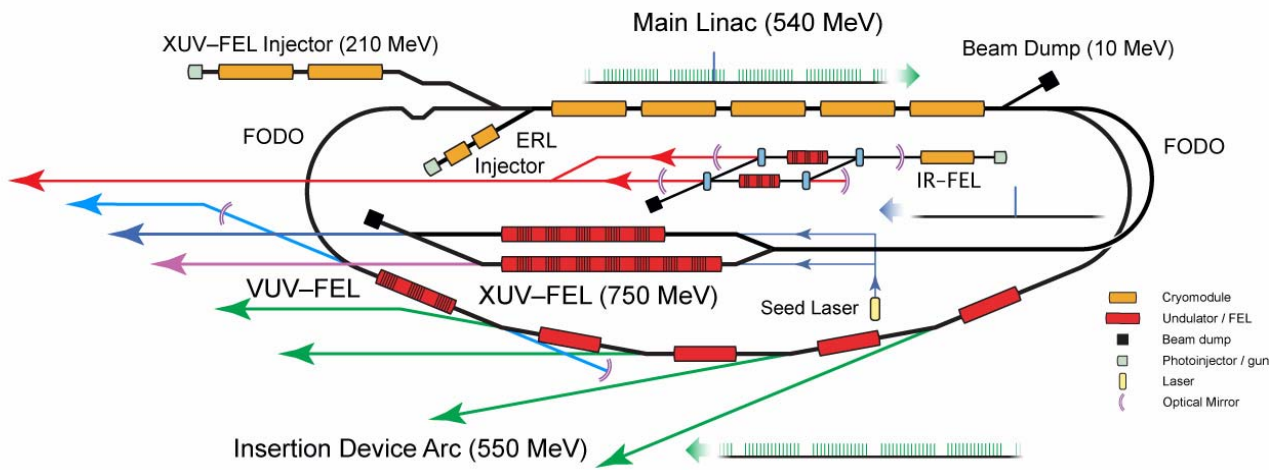

Figure 1: Schematic layout of the proposed 4GLS facility showing principal accelerator sections.

Table 1: Output electron bunch parameters of the 4GLS facility branches.

| | XUV-FEL | ERL (100 mA) | ERL (VUV-FEL) | IR-FEL |
|---|---|---|---|---|
| Energy | 750 MeV | 550 MeV | | 25-60 MeV |
| Bunch Rate | 1 kHZ | 1.3 GHz | 4.33 MHz | 13 MHz |
| Bunch Charge | 1 nC | 77 pC | | 200 pC |
| Normalised Emittance | | 2 mm-mrad | | 5-10 mm-mrad |
| Projected Energy Spread | | 0.1 % | | 0.1% (60 MeV) |
| r.m.s. Bunch Length | < 270 fs | 100-500 fs | 100 fs | 1-10 ps |
| Average Beam Power | 1 kW | 55 MW | 180 kW | 156 kW (60 MeV) |

## ERL FEL POWER LIMIT

The VUV-FEL is a regenerative amplifier and achieves saturation in approximately ten to fifteen passes [7]; at saturation, the final electron bunch energy spread is dominated by the lasing itself, and GENESIS 1.3 steady-state simulations [8] predict a full spread of approximately 1.2 %. A simple model of FEL lasing allows us to examine the scaling of laser power with energy spread at the dump. It can be shown [9] that, for small initial electron beam energy spread, the energy spread after lasing is $\sigma_{FEL} \sim \rho A^2$, where $\rho$ is the FEL Pierce parameter and $A \sim 1$ is the scaled field amplitude. The mean relative energy loss has the same value, so that

$$\sigma_{FEL} \sim \frac{\langle \Delta E \rangle}{E_{FEL}} \sim \rho A^2.$$

In the case that the absolute energy spread is conserved in an ERL, the FEL average power must be less than the beam power incident on the dump, since the relative energy spread at the dump must be less than 100 %. The photon energy per pulse is

$$E_p = n_e \langle \Delta E \rangle << E_r,$$

where $E_r$ is the average dump energy. For bunch frequency $f$, we have

$$P_{FEL} = f E_p, \quad P_{dump} = n_e f E_r,$$

therefore

$$P_{FEL} << P_{dump}.$$

For 4GLS, since the maximum dump power is simply the beam power at 10 MeV – 1 MW, the average FEL power is limited to around 100 kW whatever the bunch frequency. Power limits on the VUV-FEL mirrors will impose a much lower limit than that. However, we can express the power limit as a limit from the energy spread

$$P_{FEL} < \sigma_r \frac{q_r f E_r}{e},$$

where $\sigma_r$ is the relative energy spread at the dump. Conservatively, limiting the final energy spread to 10 % gives an average power limit for the 4GLS VUV-FEL of 300 W.

Although the absolute energy spread can be changed somewhat by the deceleration process, we still have a scaling of the energy spread at the dump with extracted laser pulse energy. A 1D simulation of the 4GLS lasing confirms this [10]; as the energy spread from lasing increases so does the energy spread at the dump.

## SEXTUPOLE LINEARISATION

The present optical configuration will use sextupoles in the outward FODO arcs to perform linearisation of the RF curvature from the main linac. In principle, either $3^{rd}$ harmonic RF or sextupoles ($T_{566}$) may perform linearisation. For a long initial bunch length sextupole linearisation is not as effective as using a $3^{rd}$ harmonic cavity. An optimisation has been done for the 4GLS ERL case which shows that sextupole linearisation is effective if the bunch length at the exit of the high-current injector accelerating modules is less than 3 ps. Simulations of the high-current gun show that this bunch length is achievable [5].

## SYNCHRONISATION OF THE XUV-FEL

Of critical importance to any seeded FEL is the correct timing of the seed pulses and the electron bunches to ensure their coincidence at the start of the FEL interaction region. For the 4GLS XUV-FEL [6], it is desirable to lock the seed pulse and Gaussian electron bunches together with a jitter < 50 fs. A 1D model has been used to track the electrons from the output of the gun through the accelerating and dispersive sections. It allows the dependence of the bunch arrival time on the individual RF phases and amplitudes to be determined. Table 2 summarises the jitter contributions for the various XUV-FEL synchronisation sub-systems [6].

Table 2: Contributions to timing jitter between seed photon pulses and bunches at the XUV-FEL entrance. Gaussian bunches are assumed.

| Source | Timing Jitter [fs] |
|---|---|
| Master Clock (0.1 – 1 kHz) | 20 |
| Distribution to seed laser oscillator | 15 |
| Seed laser oscillator | 20 |
| Amplifier & photon beam transport | 5 |
| Electron gun drive laser | < 1 |
| RF phase in accelerator cavities | 31 |
| RF amplitude in accelerator cavities | 145 |
| Total (assuming uncorrelated) | 152 |

It should be noted that the tabulated values assume a Gaussian current profile for the electron bunches. In reality, we expect the current profile from the XUV injector to resemble that of BESSY FEL. Their modelling predicts electron bunches with an almost `flat-top' current of 2 kA of length 730 fs [11]. In this case an acceptable temporal jitter would be slightly less than half of this value considerably relaxing the 4GLS design target of < 50 fs. The combined jitter estimate of 152 fs above would therefore easily satisfy the jitter tolerances for a BESSY-type electron bunch.

## ERL PATH LENGTH CORRECTION

In order to perform energy recovery the bunches must return to the main linac π out of RF phase with respect to the accelerating bunches. Therefore we must be able to introduce extra path length without affecting other beam parameters through the machine. To allow flexibility in tuning and operation, and to allow the possibility of a second accelerating pass in a future upgrade, we require a full wavelength (23 cm) of adjustment. A modular system decoupled from the rest of the accelerator is both more compact and will be simpler to operate.

We propose a novel system to introduce a continuously-variable path-length difference without introducing any variation of longitudinal dispersion. Our approach combines a magnetic chicane (large positive $R_{56}$ in our convention) and two, physically moving, non-dispersive doglegs (small negative $R_{56}$) [12]. The moving doglegs are coupled by a set of bellows that expand to introduce most of the required extra path length (Figure 2). The small $R_{56}$ induced by the changed position of the dogleg is cancelled by a small magnetic adjustment in the chicane; this ensures $R_{56}$ remains constant as required.

In principle, the system can perform the specific task of path length correction while leaving all other parameters in the accelerator unchanged. We relax this somewhat to use this section to compensate for $R_{56}$ generated in the insertion device arc and VUV-FEL, thus ensuring correct bunch decompression for re-entry into the main linac in such a way that the energy spread at the dump is minimised. The system is located just prior to re-entry of the beam into the main linac (Figure 3). Including the decompression chicane, it is only around 10 m long; the moving part 6 m. The horizontal displacement can be up to 900mm. This should be compared to a total moving length of around 40m if this task was performed mechanically in one of the main ERL arcs.

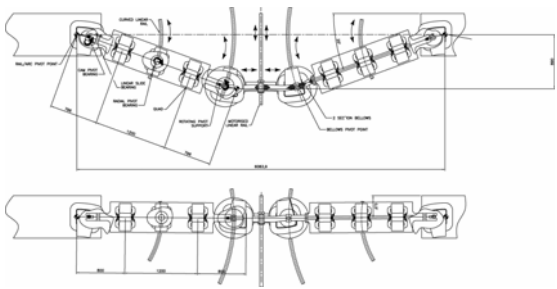

Figure 2: The moving dogleg in its maximal (top) and minimal (bottom) displacement configurations.

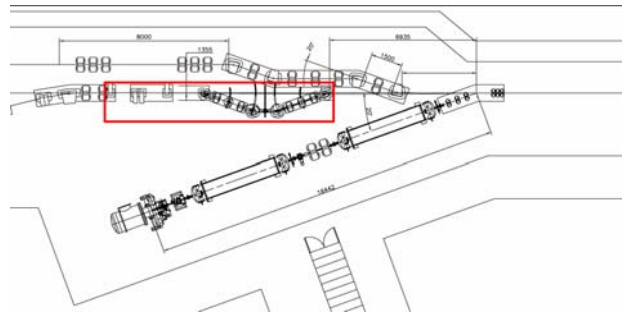

Figure 3: Outlined is placement of path correction / decompression system just prior to re-entry into the main linac. Above it is the XUV injector, below is the high-current ERL injector. The main linac is to the right.